\documentclass{ws-ijgmmp}

\newcommand{\vv}{``}

\begin{document}

\markboth{C. Branchina, V. Branchina, F. Contino and A. Pernace}
{Does the Cosmological Constant really indicate the existence of a Dark Dimension?}

%
\catchline{}{}{}{}{}
%

\title{DOES THE COSMOLOGICAL CONSTANT REALLY INDICATE THE EXISTENCE OF A DARK DIMENSION?}

\author{CARLO BRANCHINA$^{1\,2\,\dagger}$, VINCENZO BRANCHINA$^{3\,}$\footnote{Corresponding author}\,\,\,, FILIPPO CONTINO$^{3\,\ddagger}$, ARCANGELO PERNACE$^{3\,\star}$}

\address{
	${}^1$Department of Physics, Chung-Ang University, \\
	Seoul 06974, Korea \\
	${}^2$Department of Physics, University of Calabria, and INFN-Cosenza\\
	Arcavacata di Rende, I-87036, Cosenza, Italy \\
	${}^3$Department of Physics, University of Catania, and INFN-Catania\\
	Via Santa Sofia 64, I-95123 
	Catania, Italy\\
	\email{${}^\dagger$carlo.branchina@unical.it} 
	\email{${}^*$vincenzo.branchina@ct.infn.it} 
	\email{${}^\ddagger$filippo.contino@ct.infn.it}
	\email{${}^\star$arcangelo.pernace@ct.infn.it} 
}

\maketitle

\begin{history}
\received{(Day Month Year)}
\revised{(Day Month Year)}
\end{history}

\begin{abstract}
According to the \vv dark dimension" (DD) scenario, we might live in a universe with a single compact extra dimension, whose mesoscopic size is dictated by the measured value of the cosmological constant. This scenario is based on swampland conjectures, that lead to the relation $\rho_{\rm swamp}\sim m_{_{\rm KK}}^4$ between the vacuum energy $\rho_{\rm swamp}$ and the size of the extra dimension $m_{_{\rm KK}}^{-1}$ ($m_{_{\rm KK}}$ is the mass scale of a Kaluza-Klein tower), and on the corresponding result $\rho_{_{\rm EFT}}$ from the EFT limit.
We show that $\rho_{_{\rm EFT}}$ contains previously missed UV-sensitive terms, whose presence invalidates the widely spread belief (based on existing literature) that the calculation gives automatically the finite result $\rho_{_{\rm EFT}}\sim m_{_{\rm KK}}^4$ (with no need for fine-tuning). This renders the matching between $\rho_{\rm swamp}$ and $\rho_{_{\rm EFT}}$ a non-trivial issue. We then comment on the necessity to find a mechanism that implements the suppression of the aforementioned UV-sensitive terms. This should finally allow to frame the DD scenario in a self-consistent framework, also in view of its several phenomenological applications based on EFT calculations.
\end{abstract}


\section{Introduction}
Theories with large extra 
dimensions were extensively explored in the nineties in  search for a solution to the electroweak naturalness/hierarchy problem\,\cite{Antoniadis:1990ew,Arkani-Hamed:1998jmv,Antoniadis:1998ig}. 
A recent surge of interest towards the physics of $5$D effective field theories (EFTs) with one compact extra dimension of mesoscopic size has followed the dark dimension (DD) proposal, which, following arguments based on swampland conjectures\,\cite{Montero:2022prj}, and born in a string framework, suggests the existence of a single extra 
dimension of $\mu {\rm m}$ size. 
A generic feature of string theories is the existence of  towers with an infinite number of states, whose masses are given in terms of a scale $\mu_{{tow}}$. According to the swampland distance  conjecture\,\cite{Ooguri:2006in}, at large distance in the moduli field space $\phi$ one of the tower scales becomes exponentially small, $\mu_{tow} \sim e^{-\alpha |\phi|}$ ($\alpha$ positive $\mathcal O(1)$ constant), and the DD proposal is related to this asymptotic regime.
As stressed in\,\cite{Montero:2022prj}, in these regions 
only two cases seem to arise in string compactifications: a tower of string excitation modes or a tower of Kaluza-Klein (KK) states, i.e.\,\,$\mu_{tow} \sim M_s, m_{_{\rm KK}}$ (Emergent String Conjecture \cite{Lee:2019wij,Lee:2019xtm}). In general, in $d$ spacetime non-compact dimensions an infinite tower of states contributes to the vacuum energy $\rho_d$ an amount\, $\rho_d \sim \mu_{tow}^d$\,\cite{Polchinski:1985zf,Dienes:1995pm}. A similar result seems to hold even in the framework of 
higher dimensional field theories with compact extra dimensions. This is for instance the case for supersymmetric theories with Scherk-Schwarz or brane-localized SUSY breaking
\cite{Barbieri:2000vh,Arkani-Hamed:2001jyj}. Only KK modes are present and the usual calculation gives $\rho_d \sim m_{_{\rm KK}}^d$. 

Going back to the string (quantum gravity) framework, when the distance conjecture is implemented in AdS spaces\,\cite{Lust:2019zwm} 
\begin{equation}\label{AdSconj}
	\mu_{tow} \sim |\widetilde \Lambda_{\rm cc}|^\gamma\,,
\end{equation}
where $\widetilde\Lambda_{\rm cc}$ is the cosmological constant times the squared Planck mass $M_P^2$. Even though there is much wider support in AdS, the conjecture is nonetheless extended also to dS spaces, where it forms the basis for the dark dimension proposal\,\cite{Montero:2022prj}. 
Restricting to the $d=4$ case, the one-loop string  calculation of  $\rho_4$ gives
\begin{equation}\label{rho}
	\rho_4 \sim \mu_{tow}^4\,.
\end{equation}
The authors of \cite{Montero:2022prj} note that higher loops might only contribute with higher powers of $\mu_{tow}$, so that (barring cancellation of the $\mu_{tow}^4$ term) the comparison of \eqref{rho} with \eqref{AdSconj} gives\footnote{They also refer to \cite{Rudelius:2021oaz,Castellano:2021mmx} to further support the bound \eqref{bound}.}
\begin{equation}\label{bound}
	\gamma \geq \frac 14\,.
\end{equation}
They assume \eqref{bound} as starting point for their proposal. 
Moreover, observing that the  experimental bounds on possible violations of the ${1}/{r^2}$ Newton's law\,\cite{Lee:2020zjt} give $\mu_{tow} \geq 6.6$ meV,  and that the energy scale associated to the measured value of $\widetilde \Lambda_{\rm cc}$\,\cite{Planck:2018vyg} is of the same order, $\widetilde \Lambda_{\rm cc}^{1/4} \sim 2.31$ meV, they infer that \eqref{AdSconj} is saturated with $\gamma=1/4$, 
and accordingly the \vv experimental value'' of $\mu_{tow}$ is
\begin{equation}
	\mu_{tow}^{exp}\sim 2.31\, {\rm meV}\, 
	\label{DD2}
\end{equation}
(order the neutrino scale).
Finally they observe that, although it is in principle possible that $\mu_{tow} = M_s$,  
Eq.\,\eqref{DD2} indicates that this option  is \vv ruled out by experiments'' 
since we know that physics above the neutrino scale is well  described by effective field theories, and no sign of string excitations is observed at these scales. They then  conclude that the only possibility left is an \vv EFT decompactification scenario", with  a Kaluza-Klein mass $m_{_{\rm KK}}\sim \mu_{tow}^{exp}\sim 2.31\, {\rm meV}$.

This conclusion takes us from the string theory realm to the EFT terrain, and is crucial to the formulation of the DD proposal. Typically, when physics is described in terms of a string KK tower, the original string theory is replaced by the corresponding higher dimensional EFT with compact extra dimensions. A thorough analysis of this delicate step is one of the goals of the present work.

According to\,\cite{PDG}, the strongest bounds for the compactification scale $m_{_{ KK}}^{-1}$ come from the heating of neutron stars due to the surrounding cloud of trapped KK gravitons\,\cite{PDG,Hannestad:2003yd}, which yields to the upper bounds: $m_{_{\rm KK}}^{-1} < 44\, \mu{\rm m} $ for $n=1$,  $m_{_{\rm KK}}^{-1} < 1.6 \times 10^{-4}\, \mu{\rm m} $ for $n=2$, with more stringent bounds for $n>2$. The authors of \cite{Montero:2022prj} then conclude that $n\geq2$ is excluded since it is not compatible
with $m_{_{\rm KK}}\sim \widetilde \Lambda_{\rm cc}^{1/4}$, and that there should be a {\it single} extra dimension, they call it {\it dark dimension}, of size $\sim 1-100\,\mu {\rm m}$.

In string theory the finite result $\rho_d \sim \mu_{tow}^d$ arises from modular invariance, that requires to sum over the infinite tower of states. In higher dimensional field theories with compact extra dimensions such an UV-insensitive result for $\rho_d$ is obtained performing the calculation in a similar manner, i.e. summing over the infinite number of KK-states (the same is done for the calculation of the 4D Higgs effective potential and Higgs boson mass). Differently from the string theory case, however, in this EFT framework such a way of performing the calculation is less obvious to justify\,\cite{Ghilencea:2001ug}. In fact, this question was at the centre of a heated debate in the early 2000's. Several authors tried to support this way of operating with different arguments\,\cite{Contino:2001gz,Delgado:2001ex}, and  
even nowadays there are attempts at justifying it from the  string theory side\,\cite{Abel:2021tyt}. 

This issue, and more generally the question of developing a well-founded EFT approach to field theories with compact extra dimensions, was recently re-analysed in \cite{Branchina:2023rgi}, where the focus was  on the problem of the UV-(in)sensitivity of the one-loop Higgs effective potential and Higgs mass. It was shown that within the usual calculations the asymptotics of the loop momenta are mistreated, and that this results in an artificial washout of UV-sensitive terms of topological origin. The latter stem from the boundary conditions that must necessarily be given to define the theory on a multiply connected manifold. Their presence was first pointed out in\,\cite{Branchina:2023rgi}. 

In this work we show that, when a proper EFT calculation of the vacuum energy $\rho_{4}$ is performed, UV-sensitive terms arise (for an introduction to the cosmological constant problem see for instance\,\cite{Weinberg:1988cp, Carroll:2000fy, Copeland:2006wr}). 
Moreover we discuss how the EFT logic can and must be consistently applied to theories with compact extra dimensions, although it has been recently argued that no controlled approximation can be obtained cutting a KK tower at a finite value\,\cite{Burgess:2023pnk}. 

\section{Vacuum energy }
For concreteness, in the following we stick to the case (sufficient for our scopes) of a $5$D EFT coupled to gravity, where the compact space dimension is in the shape of a circle of radius $R$. We take the $5$D action to be 
\begin{equation}
	\mathcal{S}^{(4+1)}=\mathcal{S}_{\rm grav}^{(4+1)}+\mathcal{S}_{\rm{matter}}^{(4+1)},
	\label{free scalar action}
\end{equation}
where
\begin{equation}
	\label{action}
	\mathcal{S}_{\rm grav}^{(4+1)}=\frac{1}{2\hat{\kappa}^2}\int d^4xdz \sqrt{\hat{g}}\,\left(\hat{\mathcal R}-2\hat\Lambda_{cc}\right)
\end{equation}
is the Einstein-Hilbert action in $4+1$ dimensions and $\mathcal S^{(4+1)}_{\rm matter}$ the matter action that contains the bosonic and fermionic fields of the theory. We indicate with $x$ the $4$D coordinates and with $z$ the coordinate along the compact dimension. 
After integration over $z$, the $5$D metric (we use the $(+,-,-,-,-)$ signature)
\begin{equation}
	\label{metric}
	\hat{g}_{_{MN}}=\begin{pmatrix}
		e^{2\alpha\phi}g_{\mu \nu}-e^{2\beta \phi}A_{\mu}A_{\nu} & e^{2\beta \phi}A_{\mu} \\ 
		e^{2\beta \phi}A_{\nu} & -e^{2\beta \phi}
	\end{pmatrix}
\end{equation}
leads to the $4$D action\,\cite{Benakli:2022shq}
\begin{align}
	\label{grav4D}
	&\mathcal{S}^{(4)}_{\rm grav}=\frac{1}{2\kappa^2}\int \mathrm{d}^4x\,\sqrt{-g}\,\left[ \mathcal R-2e^{2\alpha\phi}\hat\Lambda_{cc}+{2}\alpha \Box \phi+\frac{(\partial \phi)^2}{2}-\frac{e^{-6\alpha\phi}}{4}F^2\right],
\end{align}
where the $4$D constant $\kappa$ is related to the $5$D $\hat{\kappa}$ by 
\begin{equation}
	\label{relation M_P different dimensions}
	\kappa^2=\frac{\hat{\kappa}^2}{2\pi R}.    
\end{equation}
The constants $\alpha$ and $\beta$ satisfy the relation
\begin{equation}
	\label{relation between alpha and beta}
	2\alpha+\beta=0
\end{equation}
and the canonical radion kinetic term fixes
\begin{equation}
	\label{alpha}
	\alpha=\frac{1}{\sqrt{12}}.
\end{equation}
For completeness we recall that the Newton constant $\kappa$ can be used to write \eqref{grav4D} in terms of dimensionful $\phi$ and $A_\mu$ fields through the redefinition
\begin{equation}
	\label{physical fields}
	{\phi}\to \frac{\phi}{\sqrt{2}\kappa} \,\,\,\,, \,\,\,\, {A_{\mu}}\to \frac{A_{\mu}}{\sqrt{2}\kappa}.
\end{equation}

Let us consider the case of a complex $5$D scalar field $\hat \Phi$ with action
{\small \begin{equation}
		\label{scalar action}
		\mathcal{S}_{\Phi}^{(4+1)}=\int d^4x dz\,\,\sqrt{\hat{g}}\left(\hat{g}^{MN}\partial_M \hat\Phi^* \partial_N \hat\Phi-m^2|\hat\Phi|^2\right),
\end{equation}}

\noindent
that along the compact dimension obeys the non-trivial boundary condition 	
\begin{equation}\label{BC}
	\hat\Phi(x,z+2\pi R)=e^{2\pi i \delta}\,\hat\Phi(x,z),
\end{equation}
where $\delta$ is a generic phase. Defining now ($[\delta]$ is the integer part of $\delta$)
\begin{equation}\label{q}
	q\equiv \delta-[\delta]\,,
\end{equation}
\eqref{BC} can be rewritten as
\begin{equation}\label{BCq}
	\hat\Phi(x,z+2\pi R)=e^{2\pi i q}\,\hat\Phi(x,z)\,.
\end{equation}
It is then clear that the physical parameter to which we have access is not $\delta$, but rather $q$. In other words, to implement the non-trivial boundary condition we have to refer to \eqref{BCq} rather than to \eqref{BC}. Consequently, in physical quantities $q$ rather than $\delta$ will appear.

For the corresponding $4$D action we have
\begin{align}
	\label{phi4D}
	&\mathcal S^{(4)}_\Phi=\int d^4x \sqrt{-g}\,\,\sum_{n}\left[\left|D\varphi_{n}\right|^2-\left(e^{2\alpha\phi} m^2+e^{6\alpha\phi}\frac{(n+q)^2}{R^2}\right)\left|\varphi{_n}\right|^2 \right],
\end{align}
where 
\begin{equation}
	D_\mu\equiv \partial_\mu-i\left(\frac{n+q}{R}\right)A_\mu
\end{equation}
and $\varphi_n(x)$ are the KK modes of $\hat \Phi(x,z)$.  
Taking a constant background for the radion (that for notational simplicity we continue to call $\phi$) and the trivial background for $A_\mu$, 
\begin{equation}
	\label{bmetric}
	\hat{g}^0_{_{MN}}=\begin{pmatrix}
		e^{2\alpha\phi}\eta_{\mu \nu} & 0 \\ 
		0 & -e^{2\beta \phi}
	\end{pmatrix},
\end{equation}
from \eqref{phi4D} we can define the $\phi$-dependent radius $R_\phi\equiv R\, e^{-3\alpha\phi}$ ($R\, e^{(\beta-\alpha)\phi}$ before using \eqref{relation between alpha and beta}) and the $\phi$-dependent mass $m^2_\phi\equiv m^2 e^{2\alpha\phi}$, so that the KK masses are 
\begin{equation}
	\label{KK masses}
	m_n^2\equiv m^2_\phi+\frac{(n+q)^2}{R^2_\phi}.
\end{equation}

Going to Euclidean space and considering the general case, the one-loop contribution to 
the $4$D vacuum energy $\rho_4$ of a single bosonic or fermionic tower of mass $m$ and boundary charge $q$ is then 
\begin{equation}
	\rho_{4}\sim
	(-1)^{\delta_{if}}\sum_n\int\frac{d^4p}{(2\pi)^4}\log\frac{p^2+\frac{(n+q)^2}{ R_\phi^2}+ m_\phi^2}{\mu^2},
	\label{onel vacuum energy}
\end{equation}
where $\mu$ is a subtraction scale, and $i=b,f$ for bosons and fermions respectively. 

The right hand side of \eqref{onel vacuum energy} is calculated according to different strategies. One of them consists in performing the sum over $n$ all the way up to infinity, and the integral in $d^4p$ with the help\,\footnote{We note that the introduction of $\Lambda$ is necessary, otherwise each of the integrals in \eqref{onel vacuum energy} would be divergent.} of a cutoff $\Lambda$ \cite{Barbieri:2000vh,Arkani-Hamed:2001jyj,Delgado:1998qr}. Other methods are related to the implementation of the proper time \cite{Antoniadis:2001cv}, Pauli-Villars \cite{Contino:2001gz}, thick brane \cite{Delgado:2001ex}, and dimensional regularizations \cite{Ghilencea:2005vm}. They all give the same result. For the time being we focus on the first of them; we will comment on the others later. 
Crucial in getting the UV-insensitive result $\rho_{4} \sim R_\phi^{-4}=m_{_{\rm KK}}^4$ in \cite{Barbieri:2000vh,Arkani-Hamed:2001jyj,Delgado:1998qr} is that $n$ is sent to infinity while $\Lambda$ is kept fixed\footnote{Strictly speaking, this is true only in a supersymmetric theory. Moreover, sending $\Lambda \to \infty$ at the end of the calculation is harmless, as the only $\Lambda$-dependent terms all vanish in this limit.}.
As mentioned in the Introduction, however, this way of performing the calculation mistreats the  asymptotics of the $5$D loop momentum of the original theory. In fact, $n/R$ is the fifth component $p_5$ of the 5D loop momentum $\hat p\equiv (p,n/R)$. Sending $p_5 \to \infty$ while keeping $\Lambda$ fixed means that in the loop corrections we are (improperly) including first  the asymptotics of the fifth  component ($n/R$) of the momentum and only later those of the other four components $p_1$, $p_2$, $p_3$ and $p_4$.   
As shown in \cite{Branchina:2023rgi}, however, a necessary and physical requirement, overlooked in previous literature, is that the asymptotics of all the five components of $\hat p$ have to be treated on an equal footing. 
This can be realized considering in \eqref{onel vacuum energy} a 5D cutoff, $\hat g^{^{MN}}_{_0}\hat p_{_M}\hat p_{_N}=e^{-2\alpha\phi}p^2+e^{-2\beta\phi}n^2/R^2\equiv \widetilde p^{\,2}+n^2/\widetilde R^2\le \Lambda^2$, or equivalently through the insertion of a multiplicative smooth cutoff function $e^{-(\widetilde p^2+n^2/\widetilde R^2)/\Lambda^2}$ ($\hat g^{^{MN}}_{_0}$ is the inverse of the Euclidean flat background $5$D metric in \eqref{bmetric}).
Sticking to the first of these two options, defining $\Lambda_\phi\equiv \Lambda e^{\alpha\phi}$, and performing the integration over $p$, we get (below we write only the contribution of a bosonic tower)
\begin{align}
	&\rho_{4}=\frac{1}{64\pi^2}\sum_{n=-[R_\phi\Lambda_\phi]}^{[R_\phi\Lambda_\phi]}\Bigg\{\left(\Lambda_\phi ^2-\frac{n^2}{R^2_\phi}\right) \left(m^2_\phi+\left(\frac{n+q}{R_\phi}\right)^2\right) \nonumber\\
	&+\left(\Lambda_\phi^2-\frac{n^2}{R_\phi^2}\right)^2 \log \frac{\Lambda_\phi ^2+m_\phi^2-\frac{n^2}{R_\phi^2}+\left(\frac{n+q}{R_\phi}\right)^2}{\mu ^2}\nonumber \\
	&+\left(m^2_\phi+\left(\frac{n+q}{R_\phi}\right)^2\right)^2 \log \frac{m^2_\phi+\left(\frac{n+q}{R_\phi}\right)^2}{\Lambda_\phi ^2+m^2_\phi-\frac{n^2}{R_\phi^2}+\left(\frac{n+q}{R_\phi}\right)^2}\nonumber \\
	&-\frac12\left(\Lambda_\phi ^2-\frac{n^2}{R_\phi^2}\right)^2\Bigg\}\equiv \sum_{n=-[R_\phi\Lambda_\phi]}^{[R_\phi\Lambda_\phi]} F(n),
\end{align}
where the brackets $[...]$ indicate \vv integer part'' (to simplify the notation, but without loss of generality,  we take $\Lambda$ such that $R_\phi\Lambda_\phi$ is an integer). The sum can be performed using the Euler-McLaurin (EML) formula, 
\begin{align}\label{EML}
	&\rho_4=\int_{-R_\phi\Lambda_\phi}^{R_\phi\Lambda_\phi} dx\, F(x)+\frac{F(R_\phi\Lambda_\phi)+F(-R_\phi\Lambda_\phi)}{2}\\
	& \nonumber +\sum_{j=1}^{r}\frac{B_{2j}}{(2j)!}\left(F^{(2j-1)}(R_\phi\Lambda_\phi)-F^{(2j-1)}(-R_\phi\Lambda_\phi)\right)+R_{2r},
\end{align} 
where $r$ is an integer, $B_n$ are the Bernoulli numbers, and the rest $R_{2r}$ is given by
\begin{align}\label{resto}
	R_{2r}&=\sum_{k= r+1}^{\infty}\frac{B_{2j}}{(2j)!}\left(F^{(2j-1)}(R_\phi\Lambda_\phi)-F^{(2j-1)}(-R_\phi\Lambda_\phi)\right)\nonumber\\
	&=\frac{(-1)^{2r+1}}{(2r)!}\int_{-R_\phi\Lambda_\phi}^{R_\phi\Lambda_\phi}dx\,F^{(2r)}(x)B_{2r}(x-[x]),
\end{align}
with $B_n(x)$ the Bernoulli polynomials. Expanding for $m_\phi/\Lambda_\phi$, $q/\Lambda_\phi \ll 1$, we finally get
\begin{align}
	\rho_4&=\frac{5\log \frac{\Lambda^2 e^{2\alpha\phi}}{\mu ^2}-2}{300\pi^2}e^{2\alpha\phi}R\Lambda^5  +\frac{5 m^2+3 \frac{q^2 e^{4\alpha\phi}}{R^2}}{180 \pi ^2}e^{2\alpha\phi}R\Lambda^3\nonumber\\
	&-\frac{35 m^4+14 m^2 \frac{q^2 e^{4\alpha\phi}}{R^2}+3 \frac{q^4 e^{8\alpha\phi}}{R^4}}{840 \pi ^2}e^{2\alpha\phi}R\Lambda+\frac{m^5}{60\pi} e^{2\alpha\phi}R
	\nonumber\\
	&+\frac{3 \log\frac{\Lambda^2e^{2\alpha\phi}}{\mu^2}+2}{2880 \pi^2 R^4}e^{10\alpha\phi}R\Lambda +R_4+\mathcal O(\Lambda^{-1}),
	\label{full result}
\end{align}
where the rest $R_4$ is the UV-insensitive term 	\begin{align}\label{R4}
	R_4=&-\frac{x^2 \text{Li}_3\left(r_b e^{-x}\right)+3 x \text{Li}_4\left(r_b e^{-x}\right)+3 \text{Li}_5\left(r_b e^{-x}\right)}{128 \pi^6 R^4} e^{12\alpha\phi}\nonumber\\
	&+h.c.+\frac{3 \zeta (5)}{64 \pi ^6 R^4}e^{12\alpha\phi}+\mathcal O\left(\Lambda^{-1}\right)\,
\end{align}
with
\begin{align}
	\label{symbollitium}
	r\equiv e^{2\pi i q} \qquad , \qquad x \equiv 2 \pi e^{-2\alpha\phi} R \sqrt{m^2}\,.
\end{align}
Eqs.\,\eqref{full result} and \eqref{R4} are re-written in terms of the original $R$, $\Lambda$ and $m$ (rather than $R_\phi$, $\Lambda_\phi$ and $m_\phi$) to explicitly show the $\phi$-dependence. It is worth to note that the $4$D vacuum energy $\rho_4$  is related to the corresponding $5$D one $ \rho_{4+1}$ through the relation $\rho_4=2\pi R e^{2\alpha\phi}\, \rho_{4+1}$. 

Several comments are in order. First of all we observe that, had we made the calculation in the usual way \cite{Barbieri:2000vh,Arkani-Hamed:2001jyj,Delgado:1998qr},  all the $q$-dependent UV-sensitive terms in Eq.\,\eqref{full result} (i.e.\,\,all the $q$-dependent terms except those contained in $R_4$) would be absent, while the other UV-sensitive terms are cancelled by SUSY. In fact, while the higher dimensional SUSY imposes $m_b=m_f$ for the superpartners, $q_b$ and $q_f$ are necessarily different to have a broken SUSY spectrum at low energies. With the usual calculation we would then get the well-known result $\rho_4\sim R_4^{b}-R_4^{f}\sim m_{_{\rm KK}}^4$. However, as we explain below, such a result comes from the fact that the UV-sensitive terms proportional to powers of $q$ are artificially washed out due to an improper way of treating the asymptotics of the loop momentum. 

In this respect, we now show that the interpretation of the $5$D theory as a $4$D one with an {\it infinite} tower of states (if pushed too far) is misleading. Within this interpretational framework, in fact, it is natural to consider that the correct thing to do is to sum the infinitely many ($n \to \infty$)  Coleman-Weinberg one-loop contributions brought by each of the towers. Any reference to the original $5$D loop momentum $\hat p$ (and a fortiori to the physical meaning of $n$) is lost. If on the contrary we correctly focus on the dynamical origin of the KK states, and recognize them as different momentum eigenstates that appear in the Fourier expansion of the $5$D field $\hat\Phi(x,z)$, it is clear that sending $n \to \infty$ while keeping the modulus $p$ of the other four components fixed is unphysical. If our universe has compact extra dimensions, low-energy $4$D physical observables emerge from the piling up of quantum fluctuations above the compactification scale. In implementing such a dressing, it is clear that the components of the loop momenta must be treated in a consistent way, actually on an equal footing. We will further comment on this point later.  

To better read the result \eqref{full result}, we stress that it contains four kinds of terms: (i) $m$- and $q$-independent UV-sensitive terms; (ii) UV-sensitive terms that depend only on $m$; (iii) $q$-dependent UV-sensitive terms; (iv) UV-insensitive terms. As stressed above, in SUSY theories boson and fermion superpartners have the same mass $m$, while the boundary charges $q$ 
are necessarily different to trigger the Scherk-Schwarz mechanism. Therefore:\\
(a) in SUSY theories supersymmetry enforces cancellations between superpartners of all but the $q$-dependent terms in \eqref{full result}, so that for each supermultiplet the dominant contribution to $\rho_4$ is  controlled by the SUSY breaking parameter $q^2_b-q^2_f$, and is 
\begin{equation}\label{dominantsusy}
	\rho_{4}\sim\frac{(q^2_b-q^2_f)}{R^2}\, e^{6\alpha\phi}R\Lambda^3=(q^2_b-q^2_f)\, m_{_{\rm KK}}^2 R\Lambda^3; 
\end{equation}
(b) in non-supersymmetric theories, each of the higher dimensional fields (each tower in $4$D language) gives to the vacuum energy the dominant (uncancelled) contribution
\begin{align}\label{dominantnonsusy}
	\rho_{4}&\sim e^{2\alpha\phi} R\Lambda^5 \log\frac{\Lambda^2e^{2\alpha\phi}}{\mu^2}=m_{_{\rm KK}}^{2/3}R^{5/3}\Lambda^5\log\frac{(m_{_{\rm KK}}R)^{2/3}\Lambda^2}{\mu^2}.
\end{align}
Therefore, even in the light tower limit $m_{_{\rm KK}}\to 0$ (large negative values of $\phi$), by no means the UV-insensitive $R_4\sim m_{_{\rm KK}}^4$ term in \eqref{full result} can  overthrow these dominating contributions.  

The main point that emerges from our result \eqref{full result} is that $q$-dependent UV-sensitive terms \textit{always arise} when non-trivial boundary conditions on multiply-connected manifolds are realized (necessary for instance to implement the Scherk-Schwarz mechanism in SUSY theories), \textit{independently} of the size of the extra dimensions. On the contrary, the UV-insensitive terms in the rest $R_4$ originate from the discreteness of the momentum along the circle, generated in the present case by the hierarchy between the size of the $4$D box and the radius of the circle.

\section{Vacuum Energy and Dark Dimension}

As already stressed, taking a 5-dimensional supersymmetric EFT with one compact dimension in the shape of a circle of radius $R$, and performing the calculation in the usual manner, for the vacuum energy $\rho_4$ at the one-loop level we have
\begin{equation}\label{rho4}
	(\rho_4)^{\frac{1}{4}}=\Bigl(\sum_{i} R_4^{(i)}\Bigr)^{\frac{1}{4}}\sim {\cal C}\, m_{_{\rm KK}}\,,
\end{equation}
where $R_4^{(i)}$ is of the kind \eqref{R4} (see also \eqref{symbollitium}). The sum over $i$ includes all the bosonic and fermionic contributions, and $\mathcal C$ is an $\mathcal{O}(1)$ coefficient\footnote{A simple inspection of \eqref{R4} shows that, strictly speaking, \eqref{rho4} rigorously holds: (i) for massless theories (with or without compactified higher dimensional gravity); (ii) for theories with $\phi=0$ (Minkowski).}.
The DD proposal \cite{Montero:2022prj} is based on the assumption that in the asymptotic region of the string moduli space where the EFT emerges, the vacuum energy $\rho_{4}$ goes as $m_{_{\rm KK}}^4$, where $m_{_{\rm KK}}$ is a Kaluza-Klein mass (of order the neutrino scale) given by the cosmological constant. The authors argue that the EFT result \eqref{rho4} supports their proposal.

However, we have shown that \eqref{rho4} results from an improper treatment of the asymptotics of the loop momentum in the original $5$D theory, and that the correct result for $\rho_{4}$ is given by \eqref{full result} (with $q$-independent terms cancelled by SUSY). From this latter equation we see that for a supersymmetric theory with SUSY breaking parameter $q_b^2-q_f^2$ the dominant contribution to $\rho_4$ is \eqref{dominantsusy}, i.e. it goes as $m_{_{\rm KK}}^2 R\Lambda^3$. Far from being UV-finite as $m_{_{\rm KK}}^4$, this term is strongly UV-sensitive. We also see that, for a non-supersymmetric theory the term that dominates $\rho_4$ is \eqref{dominantnonsusy}: it scales with $m_{_{\rm KK}}$ as $m_{_{\rm KK}}^{2/3}\log m_{_{\rm KK}}$ and is UV-sensitive as $\Lambda^5\log\Lambda$. In both cases the expected $m_{_{\rm KK}}^4$ result is not recovered. 

Therefore, it seems that the DD proposal can hardly be considered a physical reality. One might think that a possible way to escape from such a conclusion is to admit that what in the literature is usually called EFT limit of string theory, framework in which the DD proposal is formulated, actually gives rise to a new  type of Effective Field Theory, far from what is usually intended by the community.
At present, however, there is no hint for that, and, as stressed by the authors of \cite{Montero:2022prj} themselves, we all know that around and above the neutrino scale physics is well described by the EFT paradigm in the usual and well-known sense. 

In this respect we also observe that to calculate the contribution of a KK tower to $\rho_4$, in particular to study its UV (in)sensitivity, we could resort to the species scale cutoff $\Lambda_{\rm sp}$\,\cite{Dvali:2007hz,Dvali:2007wp}, as it is sometimes done within the swampland program (see for instance \cite{Heidenreich:2017sim,Grimm:2018ohb,Palti:2019pca,vanBeest:2021lhn,Agmon:2022thq}).
Referring to our previous examples, we consider the case of a massless $5$D field, where the masses of the tower states are given by \eqref{KK masses} with $m_\phi^2=0$. The number $N$ of states with mass below $\Lambda_{\rm sp}$ is 
\begin{equation}\label{N}
	N\equiv n_{\rm max}+|n_{\rm min}|+1\,,
\end{equation}
with $n_{\rm max}$ and $n_{\rm min}$ solutions of\, $(n+q)^2/R_\phi^2=\Lambda_{\rm sp}^2$, i.e.
\begin{equation}\label{nmax}
	n_{\rm max}=R_\phi\Lambda_{\rm sp}-q\qquad;\qquad 	n_{\rm min}=-R_\phi\Lambda_{\rm sp}-q\,.
\end{equation}
The explicit calculation is performed in the Appendix.

We stress that when $q\neq 0$, cutting the sum over $n$ in \eqref{onel vacuum energy} with \eqref{N} and \eqref{nmax} and the integral over the four-momentum $p$ with $\Lambda_{\rm sp}$ {\it is not} equivalent to the introduction of a cut on the $5$D loop momentum $\hat p$. As repeatedly underlined in the present work, the latter is the (physically) correct cut to apply. From the calculations in the Appendix we see that, when the cut is imposed on the combination $(n+q)^2/R_\phi^2$ rather than\footnote{Except for the $e^{2\alpha\phi}$ rescaling factor, $n^2/R_\phi^2$ coincides with $p_5^2$, the square of the fifth component of $\hat p$.} $n^2/R_\phi^2$, an artificial washout of $q$-dependent UV-sensitive terms is operated.  
This again comes from a mistreatment of the $5$D loop momentum asymptotics. Similarly to what we have already seen, the application of the $\Lambda_{\rm sp}$ cut pushes too far the interpretation of the KK modes as massive states of the $4$D theory, losing sight of the original physical meaning of $n$.

These issues were also discussed in \cite{Branchina:2023rgi}, where it was shown in full generality that the inclusion of the boundary charge $q$ in the cut (whatever kind of cut) is at the origin of the artificial washout of the $q$-dependent UV-sensitive terms. In this respect, it is worth to stress that performing the infinite sum while keeping $\Lambda$ fixed is equivalent to include $q$ in the cut. As explained in the present work, both are physically illegitimate operations. The proper time \cite{Antoniadis:2001cv}, Pauli-Villars \cite{Contino:2001gz}, and thick brane \cite{Delgado:2001ex} regularizations all implement the insertion of $q$ in the cut over $n$, thus realizing the artificial washout mentioned above.
It is worth to point out that the use of dimensional regularization (DR), as done for instance in \cite{Burgess:2023pnk,Ghilencea:2005vm}, does not help to cope with this kind of issues. By construction, in fact, DR does not detect the full UV-sensitivity of a theory, since in this regularization power \vv divergences" are automatically cancelled (see\,\cite{Branchina:2022jqc} for a careful analysis of DR in comparison with other regularizations). We also note that DR totally masks the presence of UV-sensitive terms in odd dimensions, leading sometimes to the impression that no \vv divergences" appear in that case. 

\section {Compact dimensions and EFTs}

The result $\rho_4\sim m_{_{\rm KK}}^4$ (more generally $\rho_d\sim m_{_{\rm KK}}^d$) is sometimes used to argue for a possible general breakdown of EFT methods. On our side, referring to the original $(4+n)D$ theory (with $n$ compact dimensions), we have shown in the previous sections that the EFT approach is perfectly suited to theories with compact extra dimensions, and found for $\rho_4$ the radically different result \eqref{full result}. An interesting point of view on these issues has been recently given in \cite{Burgess:2023pnk}, where the result $\rho_4\sim m_{_{\rm KK}}^4$ is taken for granted but it is argued that it cannot be used as a signal of general departure from the EFT approach.

Their argument goes as follows.  
Consider a tower of states with mass spectrum  $m_n= f_n\, \mu_{tow}$. 
Cutting the sum at $n=N$ means that we include in the theory KK modes up to the $N$-th one, and exclude the states from the $(N+1)$-th up to infinity. In general, integrating out a (finite) set of fields to define a low energy EFT for the lighter ones is a consistent operation only if there is a large mass hierarchy between the fields included and those excluded. When a KK tower is cut, the hierarchy between the heaviest state included and the lightest one excluded is given by $f_N/f_{N+1}$. Being this ratio $\mathcal O(1)$ (except for the case $N=0$, that defines the $4$D EFT), the authors conclude that no EFT estimate with a finite number of KK states can ever be done, and that the $(4+n)D$ theory must necessarily contain the infinite tower.

The observation that an infinite tower of massive states cannot be divided into heavy and light fields to define an EFT for the latter ones is certainly true and interesting in its own right.
In our opinion, however, this 
line of reasoning does not apply to higher dimensional theories with compact extra dimensions. In fact, sticking for concreteness to a higher dimensional $5$D theory, we should keep in mind that the KK modes are momentum eigenstates of the original $5$D fields (and not an infinitely numerable set of massive $4$D fields), and that the $5$D theory from which the $4$D theory derives is an EFT itself. As for any EFT, this means that the $5$D momentum $\hat p\equiv (p, n/R)$ in the loops has to be cut at the scale where the 
theory loses its validity. The fifth component $n/R$ of $\hat p$ cannot be disentangled from the other four components, $p\equiv (p_1,p_2,p_3,p_4)$, so that the cut in the KK states results from a physically necessary requirement. No large hierarchy between included and excluded momentum modes is ever needed.

Starting from the $5$D action $\mathcal S_{\Lambda}^{(5)}$, and integrating out the modes in the range $[k,\Lambda]$, one obtains the action $\mathcal S_{k}^{(5)}$ at the lower scale $k$.  Due to the discreteness of $p_5=n/R$, the contribution from the related eigenmodes comes in a stepwise fashion. For $k<1/R$\, no such eigenmodes appear any longer,  and the RG evolution becomes effectively of $4$D type. It is {\it only in this sense}  that the $4$D theory emerges from the $5$D one, and no $4$D theory with an infinite tower can ever give an accurate description of the original $5$D theory.

\section{Summary and Conclusions}

In the present work	we analysed the recent dark dimension proposal\,\cite{Montero:2022prj}, according to which the tiny measured value of the cosmological constant might signal the presence of a single compact extra dimension of mesoscopic size (order $\mu m$ or so). 
Moving from swampland arguments, this proposal is based on the idea that the cosmological constant fixes the scale $m_{_{\rm KK}}$ of a KK tower (of order the neutrino scale), and relies on the (widely believed) result $\rho_{_{\rm EFT}}\sim m_{_{\rm KK}}^4$ for the vacuum energy of the underlying higher dimensional EFT with one compact extra dimension. 

According to\,\cite{Montero:2022prj}, both the relation $\rho_{\rm swamp}\sim m_{_{\rm KK}}^4$ (that comes from swampland conjectures formulated in an asymptotic corner of the quantum gravity landscape) and the corresponding result for the vacuum energy coming from the EFT limit are at the basis of the dark dimension scenario. 
Leaving aside problems that might arise in the string theory framework itself, in the present work we have shown that, due to the presence of previously missed UV-sensitive terms in $\rho_{_{\rm EFT}}$ (proportional to the $q$ charges), the matching between $\rho_{\rm swamp}$ and the corresponding $\rho_{_{\rm EFT}}$ is a delicate issue. In order for such a matching to be realised, a physical mechanism that implements the suppression of the aforementioned UV-sensitive terms is needed. 

In fact, contrary to what is typically done in the literature\,\cite{Contino:2001gz,Delgado:2001ex,Delgado:1998qr,Antoniadis:2001cv,Ghilencea:2005vm}, we have stressed that in the one-loop calculation of $\rho_{_{\rm EFT}}$ the five components of the $5$D loop momentum $\hat p=(p_1,p_2,p_3,p_4,n/R)$ have to be cut in a coherent manner (for simplicity we stick to the compactification on a circle of radius $R$, for which $p_5=n/R$). In the usual calculations, where the result
$\rho_{_{\rm EFT}}\sim m_{_{\rm KK}}^4$ is automatically obtained, the asymptotics of $\hat p$ are mistreated, and this produces an {\it artificial} washout of UV-sensitive terms. Actually, the finite result $\rho_{_{\rm EFT}}\sim m_{_{\rm KK}}^4$ comes from an incorrect way of implementing the piling up of the quantum fluctuations. For example, for a $5$D supersymmetric model with Scherk-Schwarz SUSY breaking, a UV sensitivity with dominant term $m_{_{\rm KK}}^{2}R\Lambda^{3}$ is generated (see \eqref{dominantsusy}), while for a non-SUSY model the dominant UV-sensitive term is $m_{_{\rm KK}}^{2/3}R^{5/3}\Lambda^5\log(m_{_{\rm KK}}^{1/3}R^{1/3}\Lambda)$ (see \eqref{dominantnonsusy}).

We have also shown that inconsistencies in the way of treating the asymptotics of the loop momenta appear even when the physical cut is implemented resorting to the species scale $\Lambda_{\rm sp}$. In fact, when non-vanishing boundary charges $q$ are present (as it is the case with the SUSY breaking Scherk-Schwarz mechanism), the straightforward exclusion of the KK masses \eqref{KK masses} above $\Lambda_{\rm sp}$ does not implement a physically acceptable cut on the fifth component $p_5$ of the $5$D loop momentum $\hat p$.
Naturally, what we have just said is not in conflict with the existence of the natural physical cutoff $\Lambda_{\rm sp}$, that originates from the coupling of $N$ species of fields with gravity, and correctly embodies the scale where gravity itself becomes strong and the EFT language can no longer be used \cite{Dvali:2007hz,Dvali:2007wp}. We rather formulate a warning on possible blind uses of $\Lambda_{\rm sp}$ in KK theories.

To conclude, the results of the present paper indicate that the recently proposed dark dimension scenario, at least the way it has been originally formulated in\,\cite{Montero:2022prj} and further implemented in several applications where the usual (but improper) way of implementing the loop calculation in EFTs are heavily used\,\cite{Anchordoqui:2022txe,Gonzalo:2022jac,Anchordoqui:2022svl,Blumenhagen:2022zzw,Anchordoqui:2023oqm,Anchordoqui:2023tln,Law-Smith:2023czn}, requires a great care when the swampland conjectures (string theory side) are confronted with the corresponding EFT limit results. We hope to come back to these issues in the near future.
\section*{Acknowledgments}
	The work of CB has been partly supported by the Basic Science Research
	Program through the National Research Foundation of
	Korea (NRF) funded by the Ministry of Education, Science and Technology (NRF-2022R1A2C2003567), and partly by the European Union – Next Generation EU
	through the research grant number P2022Z4P4B “SOPHYA - Sustainable Optimised PHYsics
	Algorithms: fundamental physics to build an advanced society” under the program PRIN 2022
	PNRR of the Italian Ministero dell’Università e Ricerca (MUR). The work of VB, FC and AP is carried out within the INFN project QGSKY.

\appendix

\section{}

In this appendix we perform the calculation of the vacuum energy $\rho_4$ using the species scale cutoff $\Lambda_{\rm sp}$. In a $4$D theory with $N$ particle states, $\Lambda_{\rm sp}=M_p/\sqrt{N}$. 
In a $5$D theory with one compact dimension the identification of $\Lambda_{\rm sp}$ is done counting the number of KK states that respect the condition $m^2_n\le \Lambda_{\rm sp}^2$. 
The inequality is saturated  when 
\begin{equation}
	m^2_\phi+\left(\frac{n+q}{R_\phi}\right)^2=\Lambda_{\rm sp}^2 \rightarrow n_{\pm}= \left[-q\pm R_\phi\sqrt{\Lambda_{\rm sp}^2-m_\phi^2}\,\right],
\end{equation}
where $n_{\pm}$ reduce to $n_{\rm max}$ and $n_{\rm min}$ of \eqref{nmax} in the text when $m_\phi=0$,	and the brackets $[...]$ indicate \vv integer part" (that in the following we neglect for simplicity). The number of states between $n_+$ and $n_-$ is 
\begin{equation}
	N=n_+ + |n_-|+1= 2 R_\phi\sqrt{\Lambda_{\rm sp}-m^2_\phi}+1
\end{equation} 
and the species scale is then obtained as
\begin{align}
	\label{solution species scale}
	&\Lambda_{\rm sp}=\frac{a}{3}+\frac{X^{1/3}}{3 \cdot 2^{1/3}}-\frac{2^{1/3} \left(3 b-a^2\right)}{3 \cdot X^{1/3}}
\end{align}
with 
\begin{align}
	X&=3 \sqrt{3} \sqrt{4 a^3 c-a^2 b^2-18 a b c+4 b^3+27 c^2}+2 a^3-9 a b+27 c
\end{align}
and 
\begin{align}
	a&=m_\phi^2+\frac{1}{4 R_\phi^2};\quad b= \frac{M_p^2}{2 R_\phi^2}; \quad c=\frac{M_p^4}{4 R_\phi^2}.
\end{align}
Expanding for $m_\phi, R_\phi^{-1}\ll M_p$, we get
\begin{equation}
	\label{relation species scale Planck}
	\Lambda_{\rm sp}^2= \frac{M_p^{4/3}}{(2R_\phi)^{2/3} }-\frac{M_p^{2/3}}{3\, (2 R_\phi^4)^{1/3} }+\frac{m^2_\phi+\frac{1}{4 R_\phi^2}}{3} +\mathcal O(M_P^{-2/3}).
\end{equation}
The first term of this expansion is the one typically referred to in the literature, where only a rough estimate of $\Lambda_{\rm sp}$ is reported (see for instance \cite{Grimm:2018ohb}).

The contribution of a bosonic (or fermionic, adding an overall minus sign) tower to the vacuum energy is
\begin{equation}
	\label{rho4 species scale}
	\rho_{4}\sim \sum_{n=n_-}^{n_+}\int^{(\Lambda_{\rm sp})}\frac{d^4p}{(2\pi)^4}\log\frac{p^2+\frac{(n+q)^2}{ R_\phi^2}+ m_\phi^2}{\mu^2},
\end{equation}
where the upper case $(\Lambda_{\rm sp})$ in the integral means that the modulus of the four-dimensional momentum is cut at $\Lambda_{\rm sp}$. Performing the integration over $p$ we find
\begin{align}
	&\rho_4=\frac{1}{64\pi^2}\sum_{n=n_-}^{n_+}\Bigg\{-\Lambda_{\rm sp}^4+2 \Lambda_{\rm sp}^2 \left(m_\phi^2+\frac{(n+q)^2}{R_\phi^2}\right)\nonumber\\
	&+2 \left(m_\phi^2+\frac{(n+q)^2}{R_\phi^2}\right)^2 \log \left(\frac{m_\phi^2+\frac{(n+q)^2}{R_\phi^2}}{\Lambda_{\rm sp}^2+m_\phi^2+\frac{(n+q)^2}{R_\phi^2}}\right)\nonumber\\
	&+2 \Lambda_{\rm sp}^4 \log \left(\frac{\Lambda_{\rm sp}^2+m_\phi^2+\frac{(n+q)^2}{R_\phi^2}}{\mu ^2}\right)\Bigg\}\equiv \sum_{n=n_-}^{n_+} G(n).
\end{align}
As in the text (see \eqref{EML}, \eqref{resto} and the text therein, where $B_{i}$ and $B_{i}(x)$ are defined), the sum can be calculated by means of the EML formula,  
\begin{align}
	\rho_4&=\int_{n_-}^{n_+}dx \,G(x) + \frac{G(n_+)+G(n_-)}{2} \\
	\nonumber 
	&+\sum_{j=1}^{r}\frac{B_{2j}}{(2j)!}\left(G^{(2j-1)}(n_+)-G^{(2j-1)}(n_-)\right)+R_{2r}
\end{align}
where 
\begin{align}
	R_{2r}&=\sum_{k= r+1}^{\infty}\frac{B_{2j}}{(2j)!}\left(G^{(2j-1)}(n_+)-G^{(2j-1)}(n_-)\right)\nonumber\\
	&=\frac{(-1)^{2r+1}}{(2r)!}\int_{n_-}^{n_+}dx\,G^{(2r)}(x)B_{2r}(x-[x]).
\end{align}
In the physically meaningful limit $m_\phi, R_\phi^{-1}\ll \Lambda_{\rm sp}$, the result for the vacuum energy is
\begin{align}
	&\rho_4=\frac{20 \log \left(\frac{4 M_p^2}{5 \mu ^3 R_\phi}\right)+12 \pi -57}{2^{-1/3}\cdot3840 \pi ^2 R_\phi^{2/3}}M_p^{10/3}+\frac{-4 \log \left(\frac{4 M_p^2}{\mu ^3 R_\phi}\right)-6 \pi +27}{2^{-2/3}\cdot2304 \pi ^2 R_\phi^{4/3}}M_p^{8/3} \nonumber \\
	&+\frac{12 \pi -35}{4608 \pi ^2 R_\phi^2}M_p^2+\frac{\left(4\,m_\phi^2 R_\phi^2+1\right) \log \left(\frac{M_p^2}{2 \mu^3 R_\phi}\right)-3 (5-4 \pi ) m_\phi^2 R_\phi^2}{1152 \pi ^2 R_\phi^2}M_p^2 \nonumber \\
	&+\frac{-20 \log \left(\frac{M_p^2}{\mu ^3 R_\phi}\right)-120 \pi +309+104 \log2}{2^{-1/3}\cdot 124416 \pi ^2 R_\phi^{8/3}}M_p^{4/3}\nonumber \\
	&+\frac{3(19-8 \pi )-4 \log \left(\frac{4 M_p^2}{\mu ^3 R_\phi}\right)}{2^{-1/3}\cdot 3456 \pi ^2 R_\phi^{8/3}}\left(m_\phi R_\phi\right)^2 M_p^{4/3}\nonumber \\
	&+\frac{525 \pi+367 \log2 -1953+35 \log \left(\frac{M_p^2}{\mu^3 R_\phi}\right)}{2^{-2/3}1866240 \pi ^2 R_\phi^{10/3}}M_p^{2/3}\nonumber
	\end{align}
	
	\begin{align}
	&+\frac{9 \log \left(\frac{M_p^2}{\mu^3 R_\phi}\right)+135 \pi -432+99 \log 2}{2^{-2/3}46656 \pi ^2 R_\phi^{10/3}}m_\phi^2 R_\phi^2M_p^{2/3}\nonumber\\
	&+ \frac{2 \log \left(\frac{M_p^2}{\mu ^3 R_\phi}\right)+3 \pi -30-14 \log2}{2^{-2/3}1728 \pi ^2 R_\phi^{10/3}}m_\phi^4 R_\phi^4M_p^{2/3}\nonumber \\
	&+\frac{61-18\pi + 40(17-6 \pi)m^2_\phi R^2_\phi + 80(33-9\pi)m^4_\phi R^4_\phi}{138240 \pi ^2 R_\phi^4} \nonumber\\
	&+\frac{m_\phi^5R_\phi}{60\pi}+R_4+\mathcal O(M_P^{-2/3}),
	\label{result species scale}
\end{align}
with $R_4$ given in \eqref{R4}, \eqref{symbollitium}. 

A few comments are in order. Limiting ourselves to the leading order relation $\Lambda_{\rm sp}\sim R_\phi^{-1/3} M_p^{2/3}$ (see \eqref{relation species scale Planck}), we observe that the powers $M_p^{10/3}$, $M_p^{2}$ and $M_p^{2/3}$ correspond to the powers $\Lambda^5, \Lambda^3$ and $\Lambda$ respectively in terms of a generic cutoff $\Lambda$. However, comparing the coefficients of $M_p^{2}$ and $M_p^{2/3}$ in \eqref{result species scale} with the corresponding coefficients of $\Lambda^3$ and $\Lambda$ in \eqref{full result}, we note that they have a different structure. Moreover, powers of $M_p$ other than those mentioned above (that do not find any correspondence in \eqref{full result}) are also present. These differences have a twofold origin: they are due both to the fact that the species scale cut is cylindrical in $5$D momentum space (see \cite{Branchina:2023rgi} for a thorough discussion on the difference between the implementation of a cylindrical and a spherical cutoff on the $5$D momentum) and to the fact that, as per \eqref{relation species scale Planck}, the relation $\Lambda_{\rm sp}\sim R_\phi^{-1/3}M_p^{2/3}$ holds only at the leading (large $M_p$) order.  

An even more important difference between the results \eqref{full result} and \eqref{result species scale} is that the latter does not contain any UV-sensitive term proportional to the boundary charge $q$. Actually \eqref{result species scale} comes from a physically illegitimate operation. In fact, rather than a cut on $p_5^2=e^{-2\beta\phi}\, n^2/R^2$, $\Lambda_{\rm sp}$ implements a cut on the \vv KK masses" $m_n^2=m_\phi^2+(n+q)^2/R_\phi^2$.
As discussed in the text, the (unphysical) introduction of the combination $n+q$ in the cutoff is at the origin of the artificial washout of the $q$-dependent UV-sensitive terms. This makes the result \eqref{result species scale}, and more generally the introduction of the species scale cutoff in higher dimensional theories with compact extra dimensions, unreliable. The species scale cut only arises as a result of a too literal interpretation of the $5$D theory in terms of a $4$D theory with towers of massive $4$D fields. These warnings do not apply to the case of a bona fide $4$D theory with a large number $N$ of fields coupled to gravity, where $\Lambda_{\rm sp}$ truly is the quantum gravity physical cutoff.    

It is also worth to point out that \eqref{result species scale} provides an example of a hard cutoff calculation where no $q$-dependent UV-sensitive terms are generated. In previous literature the opinion was widely expressed that the use of a hard cutoff was at the origin of UV-sensitive terms, that were considered as spurious \cite{Contino:2001gz,Delgado:2001ex,Barbieri:2001dm}. The above result shows that the presence of these terms is rather due to a correct treatment of the asymptotics of the loop momenta.

\end{document}